%% file: main.tex
\documentclass[conference,onecolumn]{IEEEtran}
\usepackage[utf8]{inputenc}
\usepackage{cite}
\usepackage[dvipsnames]{xcolor}
\usepackage{hyperref}
\usepackage{booktabs}
\usepackage{multirow}
\usepackage{comment}
\usepackage{amsmath}
\usepackage{amssymb}

\newcommand{\nmf}{NMF}
\newcommand{\dds}{DDS}
\newcommand{\ddsA}{DDS-1}
\newcommand{\ddsB}{DDS-2}
\newcommand{\ddsC}{DDS-3}

\title{Differentiable Dictionary Search: \\ Integrating Linear Mixing with Deep Non-Linear Modelling for Audio Source Separation}

\author{\IEEEauthorblockN{Luk\'a\v{s} Samuel Mart\'ak}
	\IEEEauthorblockA{\textit{CP JKU, LIT AI Lab} \\
		\textit{Johannes Kepler University}\\
		Linz, Austria \\
		lukas.martak@jku.at}
	\and
	\IEEEauthorblockN{Rainer Kelz}
	\IEEEauthorblockA{\textit{CP JKU} \\
		\textit{Johannes Kepler University}\\
		Linz, Austria \\
		rainer.kelz@jku.at}
	\and
	\IEEEauthorblockN{Gerhard Widmer}
	\IEEEauthorblockA{\textit{CP JKU, LIT AI Lab} \\
		\textit{Johannes Kepler University}\\
		Linz, Austria \\
		gerhard.widmer@jku.at}
}

\begin{document}

\maketitle
\thispagestyle{plain}
\pagestyle{plain}

\begin{abstract}

This paper describes several improvements to a new method for signal decomposition that we recently formulated under the name of Differentiable Dictionary Search (DDS).
The fundamental idea of DDS is to exploit a class of powerful deep invertible density estimators called normalizing flows, to model the dictionary in a linear decomposition method such as NMF, effectively creating a bijection between the space of dictionary elements and the associated probability space, allowing a differentiable search through the dictionary space, guided by the estimated densities.
As the initial formulation was a proof of concept with some practical limitations, we will present several steps towards making it scalable, hoping to improve both the computational complexity of the method and its signal decomposition capabilities.
As a testbed for experimental evaluation, we choose the task of frame-level piano transcription, where the signal is to be decomposed into sources whose activity is attributed to individual piano notes.
To highlight the impact of improved non-linear modelling of sources, we compare variants of our method to a linear overcomplete NMF baseline.
Experimental results will show that even in the absence of additional constraints, our models produce increasingly sparse and precise decompositions, according to two pertinent evaluation measures.

\end{abstract}


\input{content}

\section*{Acknowledgments}

The LIT AI Lab receives funding by the Federal State of Upper Austria.
GW’s work is supported by the European Research Council (ERC) under the EU’s Horizon 2020 research and innovation programme, grant agreement No. 101019375 ("Whither Music?")- 

\bibliographystyle{ieeetr} 
\bibliography{main}

\end{document}

%% file: content.tex
\section{\uppercase{Introduction}}
\label{sec:intro}
Decomposing a signal that was produced as a mixture of source signals is an essential task in signal analysis, with applications in audio source separation, automatic speech recognition, audio de-noising, musical instrument recognition, musical audio de-mixing, harmonic-percussive source separation, musical melody extraction, polyphonic music transcription, and many more.
In different domains, various assumptions are made about the nature of what is to be considered an elementary source signal, and different artifacts of decomposition are desirable, dictated by the respective objectives.

This has led to a multitude of conceptually different approaches to signal mixture decomposition.
\textit{Linear methods} such as Non-negative Matrix Factorization (NMF) have been widely extended, adjusted and customized with various constraints, to incorporate assumptions and inductive biases accommodating specificities of particular problems.
In contrast, more general and highly flexible \textit{non-linear models} such as, specifically, deep end-to-end neural networks have been successfully trained to perform mixture decomposition.
While the linear methods have inherent performance limitations relating to their modeling capacity, they seem to have an edge at both the interpretability of inference, and robustness of performance in the face of distribution shift between training and inference, when compared to these deep non-linear models.

In this work, we aim to advance an ongoing effort in integrating the strengths of these contrasting approaches.
The assumption of approximately linear mixing of sources is reflected in our use of additive source re-composition of NMF. This can be seen as a strong inductive bias on the algorithm searching for optimal signal decomposition, since it significantly constrains the size of available hypothesis space, as opposed to a non-linear mixing model.
However, the constituent sources often arise as a result of some complex non-linear physics. To address this, we assume non-linear sound producing processes for constituent sources, and train deep non-linear models to represent them. 

We present a new method for signal decomposition based on the concept of \textit{Differentiable Dictionary Search (\dds{})}, which was originally introduced in~\cite{DDS_EUSIPCO_2021}. The general idea of DDS is to use a trained explicit-likelihood density model to impose a distance measure on the semantically structured space of spectral activity that the relevant sound sources exhibit, and  use it to constrain the optimization process guiding the decomposition of an audio mixture.
This combines the efficiency of dictionary sharing across time inherited from the non-negative matrix factorization framework, with improved modelling capacity of deep normalizing flows used to express the acoustic diversity of individual sound sources.

The initial DDS formulation~\cite{DDS_EUSIPCO_2021} was mainly a proof of concept, still exhibiting a number of severe limitations.
In particular, lack of dictionary sharing and parameter reuse in source modelling resulted in scalability issues with respect to growing input size and number of sources.
In this paper, we present several extensions of the original model with the goal to further increase parameter reuse and data efficiency.
We address this by choosing to model all sources together with a \textit{single conditional density model}, with the source identity expressed by the variable being conditioned on, which we hope also leads to better generalization.
These adjustments are aimed at reducing time and space complexity of the decomposition algorithm, while improving performance at the task of source attribution.

We use audio recordings of polyphonic piano music with temporally aligned labels to evaluate the decomposition quality by assessing the precision of spectral activity attribution to the sources.
We compare variants of our method to an overcomplete NMF baseline, providing all of the compared methods with equivalent data for training (i.e. initialization), and iteration budget for inference (i.e. decomposition), ensuring that the conditions for comparison are fair, and that differences in performance can be attributed purely to differences in the methods' mechanics.
To highlight these differences, we assess the memory and computation requirements as a function of the size of training data.
Our experiments will show systematic improvement, in terms of source attribution, along the axis of method adjustments introduced by our advanced formulations.
We will also see that the adjustments lead to increasingly sparse decompositions, according to a standard sparsity measure.

\section{\uppercase{Related Work}}
\label{sec:related}
The model class used for modeling the spectral content of individual sources is called ``normalizing flows'', which have been introduced in \cite{tabak_2010}.
We used deep parametric density model architectures \textit{RealNVP} \cite{dinh_2017} and \textit{Glow} \cite{kingma2018glow}.
Adaptive dictionaries, or generative models, have been used for audio decomposition before.
Examples using Generative Adverserial Networks include \cite{DBLP:conf/icassp/SubakanS18, DBLP:conf/mlsp/0063KM20}, examples using (variational) autoencoders are \cite{DBLP:journals/corr/abs-1906-05912,DBLP:journals/spl/KaramatliCK19}. All these approaches have a shortcoming in their lack of explicit access to the likelihood of their samples. It is therefore somewhat difficult to prevent the generative models from producing samples ``too far away'' from the training data manifold. Although Variational Autoencoders can provide the likelihood of samples they generate, Monte Carlo methods need to be used to approximate an expectation of a lower bound on the likelihood \cite{Bando_2018}.
A very interesting line of research is discussed in \cite{DBLP:conf/icassp/ParkC08,DBLP:journals/tsp/LiutkusBR11,DBLP:conf/ica/FantinatoDREAJ17}, where Gaussian Processes are used for source separation. These models are adaptive, well grounded in probability theory, and even have an accessible, explicit likelihood, but are unfortunately extremely computationally demanding.

\section{\uppercase{Method}}
\label{sec:method}
The idea at the core of the \dds{} approach is to take the NMF framework $\mathbf{S} \approx \mathbf{W} \cdot \mathbf{H}$ and replace the dictionary $\mathbf{W}$ in a specific way.
A dictionary entry corresponding to a particular source represents the spectral profile of that source by using a deep generative density model.
The spectrogram $\mathbf{S}$ is decomposed into linear combinations of samples (likely spectral profiles) that are generated by these non-linear source density models.
Those samples are allowed to adapt to the input that is being decomposed.
The normalizing flow establishes a bijective differentiable mapping between the data space and a latent space, and allows explicit access to the likelihood of a sample as well.
This enables to (softly) constrain the likelihoods of samples in a natural way, which in turn (softly) constrains how far the samples can deviate from the training distribution.
The direction in which the samples should adapt is indirectly given by the gradient of the reconstruction error with respect to the latent code that produced a sample.
We are able to do this because the mapping between data space and latent space is fully differentiable.

To put it differently: by fitting generative density models to the training samples of isolated sources, we fit a parametrized function to estimate the data generating distribution. Since the model class we chose is a smooth, differentiable and invertible mapping between two manifolds (a so called \emph{diffeomorphism}), we can \emph{travel} on the data manifold by taking small steps in arbitrary directions of the high-likelihood region of the latent space to which the data manifold has been mapped. This enables us to \emph{search} through this \emph{differentiable dictionary} in any direction in data space we desire --- thus the name \emph{``Differentiable Dictionary Search''}. Several benefits of this approach have been demonstrated in \cite{DDS_EUSIPCO_2021}, through a set of highly controlled experiments.

In the following, we describe three different formulations of the \dds{} method, along with a semi-supervised NMF baseline, all of which will be evaluated in Section \ref{sec:results}. These three variants of \dds{} represent the iterative process towards improving both scalability and performance of the method.

\subsection{The Linear Baseline: \nmf{}}
We will denote the log-magnitude spectrogram of an audio snippet as $\mathbf{S} \in \mathbb{R}_+^{D \times T}$, with $D$ the spectral resolution (number of frequency bins) and $T$ the length of the input (number of time frames).
The \nmf{} approximation is given as $\hat{\mathbf{S}} = \mathbf{W} \cdot \mathbf{H}$, where $\mathbf{W} \in \mathbb{R}_+^{D \times M}$ is the dictionary matrix, and $\mathbf{H} \in \mathbb{R}_+^{M \times T}$ is the activation matrix.

The inner dimension $M$ of the \nmf{} matrix multiplication corresponds to the number of components of the decomposition, and gives the number of columns in $\mathbf{W}$ (``dictionary entries'') as well as the number of rows in $\mathbf{H}$ (``component activations'').
In an unsupervised NMF, $M$ would be a free parameter of the method.
In our experiments, $M$ will correspond to the total number of all samples (spectral frames of isolated notes) from both training and validation sets combined, that are otherwise used for training (and selection) of the dictionary model(s) in our variants of the \dds{} method.
We simply store all of these samples in the dictionary $\mathbf{W}$ to initialize the method, and keep it fixed during the decomposition.
This is what makes our baseline a semi-supervised and over-complete variant of the NMF method.
The motivation is to enable fair comparison conditions by keeping the access to raw training data equal among the methods.

We will denote the number of sources (e.g. notes in the case of piano recordings) modeled by each method as $K$. 
After decomposition, the $M$ ``component activations'' (rows in $\mathbf{H}$) are collapsed into the $K$ rows of ``source activations'' by summing components according to their source identity.

\subsection{The Unconstrained Formulation: \ddsA{}}

This is the original formulation of the \dds{} method, as proposed in \cite{DDS_EUSIPCO_2021}.
The primary deviation of the \ddsA{} formulation from NMF (as well as the other two formulations) stems from the independent treatment of time frames in the decomposed spectrogram $\mathbf{S}$.
Each time frame is modeled separately as a linear combination of $K$ different dictionary entries. %
Each of the $K$ possible sources is modeled with a separate density model using the RealNVP architecture, structure and configuration following~\cite{DDS_EUSIPCO_2021}.

The dictionary for the $k$-th source $\mathbf{W}_k \in \mathbb{R}_+^{D \times T}$ is generated by the corresponding density model $F_{\theta_k}$. Please note, that each $F_{\theta_k}$ has its own set of parameters $\theta_k$. Transforming the corresponding set of latent codes $\mathbf{Z}_k \in \mathbb{R}^{D \times T}$ as $F_{\theta_k}^{-1}(\mathbf{Z}_k^\top)^\top = \mathbf{W}_k$ for each source makes the full dictionary $\mathbf{W}$ of shape $(D \times T \times K)$ large.
The activation matrix $\mathbf{H} \in \mathbb{R}_+^{K \times T}$ then contains one scalar for scaling each dictionary entry. 
With the row vector $\mathbf{h}_k$ denoting $k$-th source activations in $\mathbf{H}$, the input approximation is given as $\hat{\mathbf{S}} = \sum_{k=1}^{K} \mathbf{W}_k \odot \mathbf{h}_k$.

This specific setup was meaningful for the evaluation of the proof of concept implementation, where we treated the monophonic test samples independently in the highly controlled experiment of ~\cite{DDS_EUSIPCO_2021}, and had only 12 sources (1 octave) in the exemplary semantic decomposition at the end.
However, it is also the source of computational bottleneck when applying this variant to larger scale problems.

\subsection{Constraining the Dictionary Size: \ddsB{}}

As a first step to improve scalability, we re-introduce the sharing of dictionary entries across time from the NMF framework, such that the input approximation is given by the product of the dictionary and activation matrices $\hat{\mathbf{S}} = \mathbf{W} \cdot \mathbf{H}$.

We introduce a new free parameter $N$ denoting ``number of \emph{components per source}'', which specifies how many dictionary entries (columns in $\mathbf{W}$) will be used to explain activity of each source. 
This yields a new structure of the dictionary $\mathbf{W} \in \mathbb{R}_{+}^{D \times KN}$ and activation $\mathbf{H} \in \mathbb{R}_{+}^{KN \times T}$ matrices, with each of the $K$ sources represented by $N$ components.
After decomposition, the $N$ ``component activations'' (rows in $\mathbf{H}$) of each source are combined by summation, to make the $K \times T$ activation matrix $\mathbf{H}$.
Dictionary entries for each source are generated by a separate, source-specific density model $F_{\theta_k}$ using the RealNVP architecture, following \ddsA{}.
With the $N$ dictionary latent codes $\mathbf{Z}_k \in \mathbb{R}^{D \times N}$ generating a subset of the dictionary as $\mathbf{W}_k = F_{\theta_k}^{-1}(\mathbf{Z}_k^\top)^\top$ for each of the $K$ sources, the full dictionary is constructed by concatenating the generated entries of all source models: $\mathbf{W} = \left [ \mathbf{W}_1, \mathbf{W}_2, \cdots , \mathbf{W}_K \right ]$.

This imposes another regularization on the adaptation of dictionary entries (other than the likelihood penalty that is fundamental to the general \dds{} approach), as the entries now need to be re-usable across multiple spectrogram frames of the polyphonic mixture.
As a result, computational efficiency is increased along with performance, due to the reduction in potential for "overfitting" any particular frame.

\subsection{Unifying the Dictionary Model: \ddsC{}}
This formulation builds on the previous one (\ddsB{}) by additionally regularizing the source modeling for the differentiable dictionary.
Instead of a set of multiple source models, a single model $F_{\theta}$ (with a unique set of parameters $\theta$), is now trained as a \emph{conditional} density model --- allowing sampling $x \sim p_{model}(x | y)$ conditioned on the source class label $y$ --- used to model the full set of dictionary entries for all $K$ sources.
To accomplish this, we choose to use an adaptation of the architecture described in \cite{jacobsen2018excessive}, which allows for the class conditioning to be expressed explicitly in the latent mapping $\mathbf{z}$ by splitting it into a \emph{semantic} part $\mathbf{z}_s \in \mathbb{R}^K$, and a \emph{nuisance} part $\mathbf{z}_n \in \mathbb{R}^{D-K}$. The source identity is expressed by a 1-hot vector in the semantic part $\mathbf{z}_s$, while most of the intra-class variance of the source should manifest in the nuisance part $\mathbf{z}_n$.
To train the conditional generative density model, we follow the training procedure from \cite{jacobsen2018excessive} up to a small change in the independence cross-entropy ($\mathit{iCE}$) objective: we replace the softmax cross-entropy on the semantic dimensions $\mathbf{z}_s$ with a mean squared error against the 1-hot class labels. Because the softmax has no explicit inverse, the conditional generation of dictionary entries using our model's explicit inverse $F_\theta^{-1}$ would become difficult.
The full dictionary $\mathbf{W} \in \mathbb{R}_{+}^{D \times KN}$ is then generated form the corresponding set of latent codes $\mathbf{Z} \in \mathbb{R}^{D \times KN}$ by the multi-source dictionary model as $\mathbf{W} = F_\theta^{-1}(\mathbf{Z}^\top)^\top$, and the decomposition is given by $\hat{\mathbf{S}} = \mathbf{W} \cdot \mathbf{H}$, following \ddsB{}.

By having a single model jointly express probability densities of all sources, the set of parameters $\theta$, albeit larger than that of any individual source model in the prior \dds{} variants, is now optimized using all of the available training samples combined.
The ratio of $\frac{\text{\# trainable parameters}}{\text{\# training data}}$ thus goes down drastically, increasing both data efficiency, and utilization of parameters.

All of the \dds{} variants obtain decomposition by minimizing the reconstruction error $ \| \mathbf{S} - \hat{\mathbf{S}} \|_2$ jointly with $-\log p(\mathbf{z})$ --- the negative log-likelihoods on the dictionary entries --- amounting to a maximum likelihood estimation (MLE) term, which is further weighted by relative contributions of dictionary entries to the full activation matrix. This objective is minimized via gradient descent by alternating updates of dictionary latent codes $\mathbf{Z}$ and the activations $\mathbf{H}$.
In the specific case of \ddsC{}, only the nuisance parts of the latent code $\mathbf{z}_n$ are updated, as semantics $\mathbf{z}_s$ are held fixed.

\subsection{Computational Complexity}
\label{sec:complexity}
To highlight how the methods that we will evaluate compare in terms of asymptotic computational complexity, we provide a rough assessment of their respective costs in an analysis with several simplifying assumptions.

Since training of the \nmf{} method corresponds to simple storage of data samples into the dictionary, it has virtually no computational cost, as opposed to the \dds{} method variants, where one or many instances of a source model need to be trained.

All compared decomposition methods proceed iteratively, and the cost of one iteration stays constant, hence giving the runtime complexity of one iteration is sufficient to compare the four different approaches.
Furthermore, since the gradient-based parameter update has a cost that is a constant factor of the forward pass cost --- the inference of $\hat{\mathbf{S}}$ from these parameters --- we limit our assessment in Table \ref{tab:complexity} to reporting on two major components of single iteration: (i) the reconstruction of input $\hat{\mathbf{S}}$ from $\mathbf{W}$ and $\mathbf{H}$, and (ii) the construction of dictionary $\mathbf{W}$ from $\mathbf{Z}$, using a set of parameters $\theta$; computed by each method as described above, omitting the associated costs of the gradient computation and gradient-based parameter update.

As inference of the dictionary $\mathbf{W}$ corresponds to passing a specific batch of data ($\mathbf{Z}$) through a (set of) normalizing flow(s), we assess this by recounting the number of matrix multiplications involved, leaving out other operations of cost proportional to the input size via constant factor, which is negligible.
We denote the number of flow steps in a normalizing flow model by $U$, and the number of matrix multiplication operations involved in a single flow step\footnote{Such as the dimension shuffling operation, or the densely connected layers of the Multi-Layer Perceptron (MLP) networks used to learn the coupling functions.} by $V$.
Further assumption that those matrix multiplications are on the order of the data dimension $D$, which approximately holds for all of our \dds{} variants parametrizations among our experiments, yields us the estimates in Table \ref{tab:complexity}.

\begin{table}[t]
	\centering
	\caption{Comparing Methods by Computational Complexity of Single Iteration}
	\begin{tabular}{@{}l|l|cccc@{}}
		\toprule
		type &
		component &
		\multicolumn{4}{c}{method} \\ \midrule
		&
		&
		\multicolumn{1}{c|}{\nmf{}} &
		\multicolumn{1}{c|}{\ddsA{}} &
		\multicolumn{1}{c|}{\ddsB{}} &
		\ddsC{} \\ \midrule
		\multirow{2}{*}{time} &
		$\hat{\mathbf{S}} \leftarrow f(\mathbf{W}, \mathbf{H})$ &
		\multicolumn{1}{c|}{$\Theta (DTM)$} &
		\multicolumn{1}{c|}{$\Theta (DTK)$} &
		\multicolumn{1}{c|}{$\Theta (DTKN)$} &
		$\Theta (DTKN)$ \\
		&
		$\mathbf{W} \leftarrow f(\mathbf{Z}, \theta)$ &
		\multicolumn{1}{c|}{---} &
		\multicolumn{1}{c|}{$\Theta (TD^2) \cdot KUV$} &
		\multicolumn{1}{c|}{$\Theta (ND^2) \cdot KUV$} &
		$\Theta (KND^2) \cdot UV$ \\ \midrule
		\multirow{2}{*}{space} &
		$\hat{\mathbf{S}} \leftarrow f(\mathbf{W}, \mathbf{H})$ &
		\multicolumn{1}{c|}{$\Theta (DM + MT + DT)$} &
		\multicolumn{1}{c|}{$\Theta (DTK + TK + DT)$} &
		\multicolumn{1}{c|}{$\Theta (DKN + KNT + DT)$} &
		$\Theta (DKN + KNT + DT)$ \\
		&
		$\mathbf{W} \leftarrow f(\mathbf{Z}, \theta)$ &
		\multicolumn{1}{c|}{---} &
		\multicolumn{1}{c|}{$\Theta (TD + D^2) \cdot KUV$} &
		\multicolumn{1}{c|}{$\Theta (ND + D^2) \cdot KUV$} &
		$\Theta (KND + D^2) \cdot UV$ \\ \bottomrule
	\end{tabular}
	\label{tab:complexity}
\end{table}

First, the \nmf{} baseline has no cost of dictionary inference, as that is fixed to the training data, but this advantage comes at the cost of \emph{linear} time and space complexity of \nmf{} with respect to the \emph{size} of training data $M$, as opposed to the \emph{constant} complexity of the \dds{} variants.
This is one of the many benefits of dictionary modeling, as typically $M \gg KN$ for most practical scenarios, such as decomposition of piano recordings.
On the other hand, the cost of dictionary modeling grows \emph{quadratically} with the spectral resolution $D$ in both time and space for all \dds{} methods, as opposed to merely \emph{linear} growth in \nmf{}. 

The computational bottleneck of the initial \ddsA{} formulation can now easily be pinpointed to the obviously \emph{linear} growth of already $D$-quadratic dictionary modeling costs in terms of temporal length $T$ of the input, which is the ultimate, fastest growing parameter in any practical scenario.
The dictionary sharing with an adjustable parameter $N$ introduced in \ddsB{} (and \ddsC{}) purposefully eliminates this linear cost.
Ultimately, the step from \ddsB{} to \ddsC{} -- consolidating the dictionary into a single model -- does not save much time in the big picture, but improves the space complexity by eliminating the $K$-factor from the $D$-quadratic cost\footnote{After multiplying by $K$ in \ddsB{}, the inference of $\mathbf{W}$ costs $\Theta (KND + KD^2) \cdot UV$ as opposed to $\Theta (KND + D^2) \cdot UV$ in \ddsC{}, making it a difference of $(K-1) D^2 \cdot UV$.}, which definitely makes a difference with higher spectral resolutions $D$ and larger sets of sources $K$.

\section{\uppercase{Experimental Setup}}
\label{sec:exp_setup}

\subsection{Data and Basic Audio Signal Processing}
To assess the relevant aspects of performance with respect to varying degrees of task complexity (such as number of concurrently sounding notes, or variance in note intensities), we use the MAPS (MIDI-Aligned Piano Sounds) dataset~\cite{MAPS_emiya2009multipitch}, as its large variety of systematically structured examples allows us to study certain aspects of our method formulations in better isolation.
We will use the ISOL subset of MAPS --- samples of isolated notes --- for training.
To evaluate the decomposition performance, we will use the RAND subset of MAPS, which contains samples of randomly generated note combinations (chords) drawn from a specific range of notes, with varying degrees of polyphony and intensity (loudness).
The MAPS dataset also allows to control for timbral variance in terms of the used instrument model and its recording conditions. We evaluate behavior of the tested systems in presence of timbral distribution shift, by carefully selecting subsets of samples for training and evaluation with no overlap along the timbre-determining axes. In the following experiments, we train all systems on samples of synthetic, sample-based instruments, excluding only the Disklavier(\texttt{ENSTDk*}) samples from training. We randomly split this set into 80\% training and 20\% validation samples. Evaluation is done on samples from the Disklavier instrument in ``Close'' recording setting --- code (\texttt{ENSTDkCl}) in \cite{MAPS_emiya2009multipitch}.

The MAPS audio samples are encoded as $44.1$ [kHz] stereo WAV files with the temporally aligned ground truth in the form of MIDI files. Before computing spectrograms, we downsample the recordings to $16$ [kHz] mono waveforms, preserving spectral activity in the frequency range of $[0; 8]$ [kHz].
We analyze the audio at spectral resolution of $512$ frequency bins, given by DFT window size of $1024$, using the Hann window function, and a hop size of $512$ samples.
The temporal resolution, as determined by the combination of sampling rate and hop size, is then used to quantize the MIDI ground truth into a piano-roll with a shape matching that of the activation matrix produced by the decomposition.
After spectral analysis of different sample audio files, silent frames are discarded based on the ground truth annotations.

\subsection{Parametrization of Methods (1): Training}

We choose to evaluate on a portion of RAND subset of MAPS that uses the pitch range of the centered 5 octaves, marked by \texttt{M36-95} in~\cite{MAPS_emiya2009multipitch}, corresponding to 60 notes between and including C2 and B6, that the evaluated methods need to model as individual sources. This range is commonly used to evaluate multi-pitch algorithms.
Note that this is a 5-fold increase from the single octave (12 notes) modelled by \ddsA{} in~\cite{DDS_EUSIPCO_2021}.

\subsubsection{\nmf{}}
The dictionary $\mathbf{W}$ in \nmf{} is initialized by loading all of the samples, which are otherwise also used for both \emph{training} and \emph{validating} the normalizing flows used in our \dds{} method formulations, and storing them directly in the dictionary, giving this method a slight edge in terms of training data volume.

\subsubsection{\dds{}}
The differentiable dictionaries in both the \ddsA{} and \ddsB{} formulations, are initialized by training a set of unconditional density models (parametrized by a modified single-scale 1-dimensional RealNVP architecture, as per Section \ref{sec:method}), one for each note.
We follow~\cite{DDS_EUSIPCO_2021} closely in regard to the model architecture and hyperparameters.
In particular, the 16 flow steps are realized via affine coupling blocks \cite{dinh2014nice} with their coupling functions approximated using MLPs with 4 dense layers of 128 units with SELU \cite{DBLP:conf/nips/KlambauerUMH17} activations. The shuffling of dimensions between the coupling blocks is realized via randomly initialized, fixed permutation layers, which are easily invertible via transposition.

The differentiable dictionary in \ddsC{} formulation is initialized by training a single conditional density model (parameterized by a modified single-scale 1-dimensional Glow architecture with semantic conditioning, as per Section \ref{sec:method}). Specifically, following the Glow architecture \cite{kingma2018glow}, each of the 32 flow steps has (i) an ActNorm layer, (ii) a trainable mixing layer parameterized in its LU-decomposition, and (iii) an affine coupling parameterized as MLP with 3 dense layers of 1024 units and Leaky ReLU \cite{xu2015empirical} activations.

All density models were trained with the Adam optimizer \cite{Kingma2015} for up to 2500 epochs using mini-batch size of 512, and learning rate of $1 \cdot 10^{-3}$ with the single-source models (\ddsA{} and \ddsB{}) and $1 \cdot 10^{-5}$ with the multi-source model (\ddsC{}). After each training epoch, the model was evaluated on the validation data for subsequent model selection.

\subsection{Parametrization of Methods (2): Inference}
For all methods, the decomposition is run for a maximum of 1000 update steps, each of which allows for two independent updates of the dictionary, and activation matrix, respectively, following the procedure of alternating updates.
\nmf{} uses a step size of $1 \cdot 10^{-2}$, the \dds{} variants use $5 \cdot 10^{-3}$.
The remaining parameters and aspects of decomposition, such as learning rate reduction upon loss plateau detection or the early stopping condition, are set equally for all methods, and follow the experimental setup outlined in~\cite{DDS_EUSIPCO_2021} identically.

\subsubsection{Initial Conditions for the \nmf{} method}
Let $M$ denote total number of samples used for training and validation, and thus also the number of entries forming the dictionary $\mathbf{W}$ in \nmf{}. Then, the activation matrix $\mathbf{H}$ is initialized with $\frac{1}{M}$ everywhere.\footnote{This by no means implies that \nmf{} is allowed twice as many updates of $\mathbf{H}$ (due to having $\mathbf{W}$ fixed) as the \dds{} methods. The maximum number of updates for each parameter is equal for all methods. A single update step allows for exactly one update of $\mathbf{H}$ and one update of $\mathbf{W}$.}.

\subsubsection{Initial Conditions for \dds{} variants}
In all variants of the \dds{} method, the latent space mapping vectors $\mathbf{Z}$ of the differentiable dictionaries are initialized to $\mathbf{0}$ vectors.
The activation matrix $\mathbf{H}$ is initialized to small, random numbers drawn from a uniform distribution $h_i \sim \mathcal{U}(0, 1)$, re-scaled to a factor of $\frac{E}{\sqrt{K}}$ for \ddsA{} and $\frac{E}{\sqrt{K \cdot N}}$ for \ddsB{} and \ddsC{}, where $E$ is the average energy of the input magnitude spectrogram $\mathbf{S}$, $K$ is number of sources, and $N$ is number of components per source.

\subsection{Post-processing}
Once the inference of activations $\mathbf{H}$ via the decomposition of input spectrogram $\mathbf{S}$ is finished, yet before calculating any performance quantifiers, we re-scale each scalar value in $\mathbf{H}$ by the norm of its associated dictionary vector from $\mathbf{W}$, the one that is ``activated'' by it.
This works slightly differently for each of the methods, and for \nmf{}, \ddsB{} and \ddsC{}, it is followed by aggregating the activity of sources over the components that represent them.
Nonetheless, the principle is the same, and this step has an equally positive effect on the performance of all methods.
The basic idea is that normalizing the dictionary entries to equal-length vectors will put the values of $\mathbf{H}$ on a rather more ``comparable'' scale of activity contribution, without burdening the optimization process of decomposition with enforcing equal norms of vectors that are being searched over.
As a result, this is a cheap way to express the activation values in $\mathbf{H}$ on a consistent, $\mathbf{W}$-independent scale.

\subsection{Evaluation Metrics}
To quantify performance of the inspected methods for a relative comparison of their properties, we calculate and report two pertinent evaluation measures, defined as follows. 

\subsubsection{Decomposition Correctness}
In order to measure the quality of decomposition, we define a metric that we call \emph{Precision of Source Attribution} (PSA). %
As per the Eq. \ref{eq:PSA}, both the activation matrix $\mathbf{H}$ and the thresholded, binary labels $\mathbf{Y}$ have the shape $[K \times T]$. The numerator in Eq. \ref{eq:PSA} quantifies the amount of activity attributed \emph{correctly}, while the denominator quantifies all of the activity attributed \emph{in total}.

\begin{equation}
\text{PSA} = \frac{\sum_{k, t} \mathbf{H} \odot \mathbf{Y}}{\sum_{k, t} \mathbf{H}}
\label{eq:PSA}
\end{equation}

Intuitively, PSA quantifies the fraction of activity in the activation matrix $\mathbf{H}$ that is attributed to those sources that are active according to the ground truth annotations.
Combined with the post-processing of $\mathbf{H}$ described above, it helps capture how well the activity in magnitude spectra is explained by the dictionary elements associated with those sources that are truly active.

\subsubsection{Decomposition Sparsity}
To quantify the sparsity of the activation matrices produced, we will report a simple measure of sparsity based on $L^0_\epsilon$~\cite{hurley2009comparing}. The standard $L^0$-sparsity simply computes fraction of the count of zeros to the total number of elements in the matrix. The $L^0_\epsilon$-sparsity also includes small, non-zero elements into that count, if they are below some $\epsilon$ threshold. We use $\epsilon=5 \cdot 10^{-2}$ to correct for noisy activity in $\mathbf{H}$, which our \nmf{} baseline is more prone to generate due to its excessive dictionary size.

\section{\uppercase{Results}}
\label{sec:results}
To examine the general trends of task difficulty and method performance on different problem configurations, we executed multiple audio decomposition experiments using audio snippets created using various parametrizations.
To highlight the relative performance of the tested methods, we first report a detailed result on a single such configuration, picked at random.
Nonetheless, the observed trends in relative performance of the compared methods seem to hold consistently across varying test configurations.

\subsection{Method Comparison}
\label{sec:perf_detailed_comparison}
The numbers reported in Table \ref{tab:methods_comparison} were obtained by running decompositions on a test snippet constructed by concatenating only segments with actual note content, excluding all silence. The segments were selected to fulfill the following criteria: (i) notes have an intensity range \texttt{I32-96} representing the more complex scenario of higher dynamic range, (ii) there are only two notes sounding at any given time. The number of individual dictionary components used for each source was set to $N=64$ for both \ddsB{} and \ddsC{}.

\begin{table}[t]
	\centering
	\caption{Improving attribution correctness.}
	\begin{tabular}{@{}c|cccc@{}}
		\toprule
		{} & \nmf{} & \ddsA{} & \ddsB{} & \ddsC{} \\ \midrule
		PSA & 0.3568 & 0.2760 & 0.4323 & 0.6475 \\ \midrule
		$L^0_{\epsilon}$ & 0.6824 & 0.7039 & 0.8121 & 0.9367 \\ \bottomrule
	\end{tabular}
	\label{tab:methods_comparison}
\end{table}

As our measure of decomposition quality PSA (Eq. \ref{eq:PSA}) indicates, each of the adjustments introduced to our formulation of the \dds{} approach led to an improvement in performance, ultimately outperforming the linear baseline by a large margin.
In particular, the introduction of dictionary sharing into \ddsB{} seems to be responsible for a $56.7\%$ relative improvement against \ddsA{}, while the additional introduction of parameter sharing in the parameterization of source model in \ddsC{} seems to have improved the decomposition quality by another $49.8\%$ with respect to \ddsB{}.
The under-performance of \ddsA{} compared to the \nmf{} baseline points towards its inefficient scaling, as discussed in \ref{sec:complexity}.

Moreover, all of the considered variants of \dds{} appear to be producing increasingly sparse decompositions, demonstrating increasing suitability to the problem domain of audio decomposition, in which mixtures are typically composed of small subsets of all potential sources. However, we observed this as an unintended positive side effect of the adjustments introduced, having designed none of them with the specific goal of encouraging sparse decompositions in mind. Using any of these methods, sparsity can still be additionally incentivized through an explicit regularization term in the decomposition objective.

\subsection{General Trends}
\label{sec:perf_general_trends}
In broader set of experiments, we varied different problem parameters and method arguments, such as spectral resolution of the input, polyphony level and dynamics range of the test samples, and the method flexibility for \ddsB{} and \ddsC{} controlled via the dictionary size argument $N$ (components per source).
While the trend of relative performance gains reported in \ref{sec:perf_detailed_comparison} appeared quite consistently, we observed several additional trends.

In general, all methods --- including the baseline --- usually achieve slightly better decomposition metrics at a larger spectral resolution of $1024$, albeit at higher computational cost. Most of the trends we observe, appear to be more pronounced at the higher spectral resolution of $1024$.
In terms of both the decomposition and reconstruction quality measures, the samples with a higher dynamic range (\texttt{I32-96}) were generally representing a harder task than those of lower dynamic range (\texttt{I60-68}), as all methods had consistently better metrics on the samples with low dynamic range.
The problem difficulty also seems to grow almost linearly with the level of polyphony, at least between \texttt{P2} and \texttt{P7}, for which we have measurements from the MAPS dataset.
When tweaking the method capacity parameter $N$, higher values in \ddsB{} produced generally smaller reconstruction error, but not necessarily more precise decomposition. On the contrary, lower $N$ seemed to produce more accurate decompositions. In contrast, with growing $N$, \ddsC{} seemed to be producing either equally good, or better decompositions than for lower values of $N$, in both decomposition quality and reconstruction error. This can be interpreted as evidence for a difference between properties of the ``bag of models'' approach to dictionary modeling of \ddsB{}, and the ``one conditional model for all sources'' approach of \ddsC{}, considering all other aspects being equal. We conjecture that the extra regularization of the single dictionary model in \ddsC{} markedly reduced the potential to ``misuse'' the spare capacity at higher $N$, as contrasted with \ddsB{} and its many less regularized models.

\section{\uppercase{Conclusion}}
\label{sec:conclusion}
We presented several adjustments to a recently proposed method for audio decomposition, improving its computation and memory costs, as well as quality of the decompositions in terms of precision of source attribution. As a positive side effect, these improvements also incentivize sparser decompositions. An investigation into these dictionary models' dynamics could reveal promising directions for further improvements.